\documentclass[CRPHYS,Unicode,manuscript]{cedram}
\usepackage{amsmath,amsfonts,amssymb,graphicx,psfrag,dsfont,slashed,bigints, bm}

\usepackage[normalem]{ulem}
\newcommand{\de}{\partial}

\newcommand{\ra}{\rangle}
\newcommand{\la}{\langle}

\newcommand{\be}{\begin{equation}}
\newcommand{\ee}{\end{equation}}
\newcommand{\ba}{\begin{eqnarray}}
\newcommand{\ea}{\end{eqnarray}}

\newcommand{\lag}{\mathcal{L}}

\newcommand{\rhot}{\tilde{\rho}}

\newcommand{\thetat}{\tilde{\theta}}
\newcommand{\mt}{\tilde{m}}

\begin{document}

\title{A diagrammatic approach to correlation functions in  superfluids}
\alttitle{Une approche diagrammatique des fonctions de corrélation dans les superfluides}

\author{\firstname{Alessia} \lastname{Biondi}\CDRorcid{0009-0000-3014-1670}}
\address{Institut Pprime, CNRS--Université de Poitiers--ISAE-ENSMA. TSA 51124, 86073 Poitiers Cedex 9, France}

\email[A. Biondi]{alessia.biondi@cnrs.fr}
\author{\firstname{Maria Luisa}  \lastname{Chiofalo}\CDRorcid{0000-0002-6992-5963}}
\address{Dipartimento di Fisica, Universit\`a di Pisa,
Polo Fibonacci, Largo B. Pontecorvo 3,  56127 Pisa, Italy}
\email[M. L. Chiofalo]{marilu.chiofalo@unipi.it}

\author{\firstname{Massimo}  \lastname{Mannarelli}\CDRorcid{0000-0003-0211-5250}\IsCorresp}
\address{INFN Laboratori Nazionali del Gran Sasso, Via G. Acitelli
22, 67100 Assergi (AQ), Italy}
\email[M. Mannarelli]{massimo.mannarelli@lngs.infn.it}

\author{\firstname{Silvia}  \lastname{Trabucco}\CDRorcid{0009-0002-7986-4735}}
\address{Gran Sasso Science Institute, Viale Francesco Crispi 7, 67100 L'Aquila, Italy}
\email[S. Trabucco]{silvia.trabucco@gssi.it}

\thanks{A.B. is supported by the CNRS Chair in Physical Hydrodynamics (reference ANR-22-CPJ2-0039-01).
This work pertains (namely, is not funded but enters in the scientific perimeter) to the French government programs ``Investissement d'Avenir'' EUR INTREE (reference ANR-18-EURE-0010) and LABEX INTERACTIFS (reference ANR-11-LABX-0017-01).\\
M.L.C. acknowledges support from the National Centre on HPC, Big Data and Quantum Computing—SPOKE 10 (Quantum Computing) and received funding from the European Union Next-GenerationEU—National Recovery and Resilience Plan (NRRP)—MISSION 4 COMPONENT 2, INVESTMENT N. 1.4—CUP N. I53C22000690001. This research has received funding from the European Union’s Digital Europe Programme DIGIQ under Grant Agreement No. 101084035. M. L. C. also acknowledges support from the project PRA2022202398 “IMAGINATION”.\\
M.L.C, M.M and S.T. acknowledge support from the INFN specific initiative NEUMATT.}

\begin{abstract}
Renaud Parentani has given a vast contribution to the development of gravitational analogue models as tools to explore various important aspects of general relativity and of quantum field theory in curved space-time. 
In these systems, two-point correlation functions are
of the utmost importance for the characterization of processes taking place close to the acoustic horizon.  In the present paper, dedicated to him, we present a study of path integral methods that allow to determine two-point correlation functions by a perturbative expansion, in a way that - beyond its generality- is especially suited to analyze these processes.  Our results apply to  non-relativistic superfluids, realizable in terrestrial experiments, as well as to relativistic superfluids, relevant for compact stellar objects.
\end{abstract} 

\begin{altabstract}
Renaud Parentani a largement contribué au développement des modèles analogues gravitationnels, utilisés comme outils pour explorer divers et importants aspects de la relativité générale et de la théorie quantique des champs en espace-temps courbe. 
Dans ces systèmes, les fonctions de corrélation à deux points revêtent une importance capitale pour la caractérisation des processus se produisant à proximité de l’horizon acoustique. 
Dans le présent article, dédié à sa mémoire, nous présentons une étude des méthodes de l'intégrale de chemin permettant de déterminer les fonctions de corrélation à deux points par un développement perturbatif, d’une manière qui - au-delà de sa généralité - est particulièrement approprié pour étudier ces processus.
Nos résultats s’appliquent tant aux superfluides non relativistes, réalisables dans des expériences en laboratoire, qu’aux superfluides relativistes, pertinents pour les objets stellaires compacts.

\end{altabstract}

\keywords{Effective field theory,  Correlation functions, Inhomogeneous superfluids.}

\maketitle

\section{Introduction}
\label{sec:intro}
Superfluidity is a macroscopic quantum effect observed in cold fluids determined by  the spontaneous breaking of a global symmetry\,\cite{Landau-stat12, Khalatnikov}. 
Below the critical temperature, the fluid's global coherence is manifested in the off-diagonal long-range order, and correlation between distant points is enhanced with respect to the normal phase, that is the case in which the temperature is above the critical value\,\cite{Penrose:1956, Hoenberg1965, Haldane:1981}. For this reason, correlation functions are the primary tool to test model predictions and compare with experimental results. They encode microscopic information on global thermodynamic quantities, on the low-energy spectrum of the fluid, and on possible quantum effects arising in the system across phase transitions. Particularly valuable examples of  superfluid systems are ultracold atoms,  which are investigated in extremely well controlled tabletop experiments. In this case it is possible to  explore different regimes at both equilibrium and out-of-equilibrium, after tuning temperature, interactions strength and range, amount of disorder, dimensionality, and the action of synthetic gauge fields\,\cite{Dalfovo:1999zz ,YagoMalo24}.

Quite generally, one can trace 
the variation of external parameters and the possible realization of different phases of matter in the correlation functions. They were the primary means to search for effects due to the emergence of a curved space-time  realized with  inhomogeneous superfluids. In particular,  considering that the low-energy excitations of the system are sound waves, a transonic flow is effectively equivalent to an emergent metric that traps  phonons: they cannot counter-propagate towards infinity\,\cite{Unruh:1994je}. 
The region where the fluid velocity equals the speed of sound is named acoustic horizon  and it is believed to emit a thermal radiation of phonons~\cite{Unruh:1994je, Brout:1995rd, Corley:1997pr,Saida:1999ap, Himemoto:2000zt, Unruh:2004zk, Barcelo:2005fc, Balbinot:2006ua, Hartley:2017fdv, Mannarelli:2020ebs, Mannarelli:2021olc, LuisaChiofalo:2022ykx, Coviello:2024vht, Trabucco:2025duz}. Such Hawking-like emission of an acoustic horizon 
produces an effective viscosity\,\cite{LuisaChiofalo:2022ykx, Trabucco:2025duz} on the fluid flow and  imprints a nontrivial signal on the density-density correlation function. The latter effect  was numerically predicted in~\cite{Carusotto}, and then confirmed in  laboratory~\cite{Steinhauer, MunozdeNova:2018fxv}.
Renaud Parentani has been one of the most influential researchers in this field, providing key insights into the theoretical understanding of different effects of the Hawking-like phonon emission on the density-density correlation function~\cite{Arteaga:2005ss, Macher:2009tw, Finazzi_2012, Michel_2016}.

In this work, we present a  path integral method to obtain the partition function of the low-energy excitations of a superfluid realized with a weakly-interacting boson gas.  This approach is motivated by applications where effective field theories are relevant and especially useful to describe the system.  Since we employ a covariant formalism,  our results could be used to characterize the properties of inhomogeneous superfluids both in terrestrial experiments and in compact stars~\cite{Shapiro-Teukolsky},  and to describe the associated phenomena, see for instance\,\cite{Poli:2023vyp}. 
After identifying the scale separation between the background fields and the corresponding excitations, we construct the gaussian low-energy effective field theory\,\cite{Georgi:annurev, Scherer:2002tk}. Then,  we study the  two-point correlation functions, as they more easily encode information on the long-range order. We provide the relevant equations to determine the correlation functions and    a  counting scheme for their recursive evaluation  in powers of momenta. Finally, we extend the approach to include inhomogeneities of the superfluid background employing techniques developed in quantum field theory\,\cite{abrikosov2012methods, Zinn-Justin:2002}. The presented results pave the way for a perturbative analytic evaluation of the effect of a sonic horizon on the two-point correlation functions.

The present paper is organized as follows. 
In Sec.\,\ref{sec:Lagrangian} we review the covariant Lagrangian  that describes the low-energy properties of a boson gas with a global $U(1)$ broken symmetry. We discuss the non-relativistic matching with the Gross-Pitaevskii model as well as the scale separation between background and excitations. In Sec.\,\ref{sec:relations} we construct the partition function for the effective field theory containing quadratic terms in the fields. In Sec.\,\ref{sec:correlators}, we evaluate the two-point correlation functions in different cases and we briefly discuss the effect  of a space-time modulated sound speed. \\
Natural units, $\hbar = c = 1$, and metric signature $(+,-,-,-)$ are used throughout this paper.

\section{Low-energy effective field theory}
\label{sec:Lagrangian}
We consider  a boson gas that is so cold that any temperature effect is negligible. For this reason, thermal fluctuations are suppressed and only quantum fluctuations are relevant. We describe this system using the mean-field approximation and we assume that the global  $U(1)$ symmetry associated to the conservation of the number of particle is spontaneously broken due to the shape of the interaction potential. According to the Goldstone theorem, the low-energy spectrum consists of a Nambu-Goldstone boson (NGB), the phonon, with a linear dispersion law\,\cite{Nielsen:1975hm}. The extension of our approach to a larger symmetry group  is straightforward.

In addition to global symmetries, space-time symmetries must be accurately taken into account\,\cite{Son:2002zn, Son_2006, Mannarelli:2008jq}. The matter background, described by a non-vanishing chemical potential, $\mu$,  explicitly breaks Lorentz boost invariance. Moreover, in the presence of  fluid flow, space rotations are explicitly broken as well. For a uniform flow, we can orient the axes in such a way that the flow four-velocity, $u_\nu$, is along the $x-$direction, so $ \boldsymbol{u} = (u,0,0)$. For simplicity, we discuss the case of vanishing transverse gradients of the fluid flow,  $\partial_y u= \partial_z u =0$, hence the system has a residual $O(2)$ symmetry corresponding to rotations around the $x-$axis. 
The relativistic Lagrangian density (hereafter the Lagrangian) of the considered  system is given by
\be
 {\mathcal L} =
(D_\nu \Phi)^* D^\nu \Phi - m^2  |\Phi|^2-\lambda |\Phi|^4\,,
\label{eq:L_general_1}
\ee
describing a complex scalar field, $\Phi$, with mass $m$ and local two-body self interaction with strength $\lambda \ll 1$. Here 
\be \label{eq:covariant_deriv}
D_\nu  = \de_\nu + i \mu u_\nu\,,
\ee
is the appropriate covariant derivative, see for instance~\cite{Son:2002zn},  determining  the  explicit breaking of the Lorentz symmetries discussed above. 
Expanding the covariant derivatives in Eq.~\eqref{eq:L_general_1},
we obtain 
\be\label{eq:L_general_2}
{\mathcal L} =\partial_\nu \Phi^*\partial^\nu \Phi  +\mu u_\nu J^\nu_{U(1)}  - (m^2-\mu^2) |\Phi|^2  - \lambda |\Phi|^4\,,
\ee
where
\be\label{eq:J_0}
J^\nu _{U(1)}= i (\Phi\partial^\nu \Phi^*- \Phi^* \partial^\nu \Phi)\,,
\ee
is the conserved current density. 
In Eq.\,\,\eqref{eq:L_general_2} we see that the chemical potential gives two distinct contributions to the Lagrangian:  the second term on the r.h.s.    produces an explicit Lorentz symmetry breaking, while the third term on the r.h.s. is an effective mass shift.  When $|\mu| = m$ the effective mass of the scalar field vanishes, signaling a symmetry  breaking  typical of second order phase transitions.

Given  the Lagrangian, 
 the generating functional  of the correlation functions is
\begin{equation}
    Z[J, J^\ast]=\! \int \! \mathcal{D}\Phi  \mathcal{D}\Phi^\ast \exp\left[ \! i \int \!  d^4x\,  \left({\mathcal L} + J^\ast\Phi  +\Phi^\ast J \right) \right],
\end{equation}
where  $J$ and $J^\ast$ are   external currents. The correlation functions can then be determined by functional derivation, for instance 
\be
\label{eq:fifi}
\la \Phi ^\ast (x_1) \Phi (x_2)\ra = - \left. \frac{1}{Z[0]} \frac{\delta^2 Z[J, J^\ast]}{\delta J(x_1) \delta J^\ast(x_2)} \right|_{J=J^\ast=0}\,, 
\ee
gives one of the  response function to   external currents. 
In order to make contact with observables accessible in laboratory, it is useful  to introduce a different representation of the  scalar field,  in terms of quantities directly linked to the fluid's density and velocity. 
\subsection{Madelung representation}
\label{sec:madelung}
In order to express the correlation functions in terms of density and velocity fields, 
we introduce the Madelung representation 
\be\label{eq:madelung}
\Phi = \frac{\rho}{\sqrt{2}} e^{i \theta/f}\,,
\ee
where both $\rho$ and $\theta$ are real scalar fields. Since  $\theta$  is a phase, it is naturally associated to the NGB arising from the spontaneous breaking of the  $U(1)$ global symmetry. 
The constant $f$ is a scaling factor  that in relativistic theories is introduced to have a scalar field with dimension of energy. For instance, in  chiral perturbation theory it corresponds to the pion decay constant, see for instance\,\cite{Scherer:2002tk},  while in second quantization for a boson system $f= \hbar$. In the following we take $f=1$ in natural units, thus $\theta$ is dimensionless. However, in Sec.\,\ref{sec:relations} we will use the freedom to rescale the phase field by an appropriate factor to simplify the notation.  

Upon substituting Eq.\,\eqref{eq:madelung}  in Eq.\,\eqref{eq:L_general_2}
we obtain the Lagrangian 
 \begin{align}
\mathcal{L}= \frac{1}{2}\de _\nu \rho \de ^\nu \rho +\frac{1}{2}\rho ^2 \de _\nu \theta \de ^\nu \theta +  \rho^2 \mu   u^\nu\de_\nu \theta -\frac{m^2  -\mu^2}{2}  \rho  ^2   - \frac{\lambda}{4}\rho^4  \,,
 \label{eq:lagr_madelung}
\end{align}
and  assuming that the system is close to equilibrium  we can expand the fields as follows: 
\be\label{eq:expansion}
\rho = \rho_0 + \rhot \qquad \theta = \theta_0 +  \thetat\,,
\ee
where $\rho_0$ and $\theta_0$ are the mean field solutions satisfying the  Euler-Lagrange equations

\begin{align}
\left.\left(\frac{\delta {\mathcal L}}{\delta \rho} - \partial_\mu \frac{\delta {\mathcal L}}{\delta \partial_\mu\rho}\right)\right|_{\rho_0, \theta_0} = 0 \,, \qquad 
\left.\partial_\mu \frac{\delta {\mathcal L}}{\delta \partial_\mu\theta}\right|_{\rho_0, \theta_0}  = 0 \,. \label{eq:EL_2}
\end{align}
The equation on the right indicates that  the current  
\be\label{eq:Jmu}
J^\nu_0 =\left.\frac{\delta {\mathcal L}}{\delta \partial_\nu\theta}\right|_{\rho_0, \theta_0}=\rho_0^2( \partial^\nu\theta_0 + \mu u^\nu)\,,
\ee
is conserved: it is the Noether current   
evaluated in the ground state. Thus,  $\partial^\mu\theta_0$ contributes to the background fluid velocity.
Assuming that the background  is homogeneous, the first equation indicates that the potential minimum is
\be
\label{eq:minimum}
 \rho_0^2  = \begin{cases} \frac{\mu ^2 -m ^2}{ \lambda} &  \text{for $|\mu| \geq m$} \\
 0 & \text{for $|\mu| < m$} \end{cases}\,,
 \ee
thus $|\mu| = m$ corresponds to the second-order quantum phase transition point. The  number density is determined by standard thermodynamic relations, and it turns out to be  
\be \label{eq:n} n=\mu \rho_0^2 \,,\ee 
therefore $\rho_0$  governs the background  density.  

In order to describe the density and velocity correlation functions, we have to rewrite the partition function in terms of the radial field fluctuation and of the phonon  field.  In detail, the density fluctuation will be proportional to $\rho_0 \rhot$, while the velocity fluctuation will be proportional to $\de^\mu \thetat$.  
From Eq.~\eqref{eq:L_general_2}, we obtain  the quadratic Lagrangian in the fluctuations  
\begin{align}
{\mathcal L}_2 = &\frac{1}2 \partial_\mu \rhot \partial^\mu \rhot -  \frac{1}{2} \mt^2 \rhot^2 + V_\nu \rhot \partial^\nu \thetat + \frac{1}{2} \rho_0^2 
\partial_\mu \thetat \partial^\mu \thetat \,,
\label{eq:lagrangian_fluctuations}
\end{align}
where for simplicity we have performed a gauge transformation in such a way that terms proportional to $\de_\mu \theta_0$ vanish,  see \cite{Trabucco:2025duz} for a  more detailed discussion. In the above equation, 
\be\label{eq:mt}
\mt = \sqrt{2 (\mu^2-m^2)} = \sqrt{  2 \lambda} \rho_0\,,
\ee
is the effective mass of the radial field fluctuation, and
\be\label{eq:Vmu} V^\nu = 2 \rho_0 \mu u^\nu\,, \ee determines the coupling between radial modes and  phonons.
The adiabatic sound speed  is given by
\be 
c_s = \sqrt{\frac{\lambda \rho_0^2}{2m^2 + 3 \lambda \rho_0^2 }}=\sqrt{\frac{\mt^2}{4m^2 + 3 \mt^2 }}  \,,
\label{eq:soundspeedm}
\ee
where the last expression is recast in terms of the effective mass of $\rhot$.

The above Lagrangian is written in a covariant form. In order to make contact with experiments, we provide the matching with the non-relativistic case in the following section.

\subsection{The non-relativistic limit and the scale separation}
The  non-relativistic limit is obtained assuming that the mass of the bosons is much larger than any other energy scale. This means that it gives the leading time dependence, therefore we write
\be
\Phi = \frac{\Phi_\text{NR}}{\sqrt{2 m}} e^{ -i m t}\,,
\label{eq:phinr}
\ee
where $\Phi_\text{NR}$ is the non-relativistic wave function. The proportionality coefficient $1/ \sqrt{2m}$ is  such that $|\Phi_\text{NR} |^2 = n$, where $n$ is the number density.
With the definition in Eq.\,\eqref{eq:phinr}, the Lagrangian in Eq.~\eqref{eq:L_general_2} for a static fluid reads
\be\label{eq:L_lambda_NR}
{\mathcal L}_\text{NR} =  i \Phi^*_\text{NR}
\de_t \Phi_\text{NR} + \Phi^*_\text{NR} \frac{\nabla^2}{2m} \Phi_\text{NR}   - \frac{\lambda}{4m^2} |\Phi_\text{NR}|^4 \,,
\ee  
where we have taken into account that in the non-relativistic limit $|\mu | \simeq m$, and we kept only the leading order in the time derivatives.
Comparing with the Lagrangian of the Gross-Pitaevskii (GP) model in vanishing external potential, \be
{\mathcal L}_\text{GP} = i \Phi^*_\text{NR}
\de_t \Phi_\text{NR} + \Phi^*_\text{NR} \frac{\nabla^2}{2m} \Phi_\text{NR}   - \frac{g}{2} |\Phi_\text{NR}|^4\,, \ee
 we have that 
\be\label{eq:g_lambda}
g = \frac{\lambda}{2 m^2}\,.
\ee
and the non-relativistic expression of the speed of sound squared is
\be
c_s^2 = \frac{n g}{m} \simeq  \frac{\mt^2}{4 m^2}\,,
\ee
where  we used Eqs.~\eqref{eq:n} and\,\eqref{eq:mt}. Alternatively, one can obtain the non-relativistic expression of the speed of sound directly from Eq.\,\eqref{eq:soundspeedm} for $\mt \ll m$. 

From the above expression it is clear that any modulation of the sound speed corresponds to a modulation of the radial field mass. This may arise as a consequence of a variation of the background density, proportional to $\rho_0^2$ or by a variation of the interaction strength $\lambda$. In this simple model these two quantities are related by Eq.\,\eqref{eq:minimum}, however we shall assume that their space dependence is different:  the background density can be kept  homogeneous by an appropriate external confinement such as an optical boxed potential~\cite{Hadzibabic}, while the speed of sound can have a non-negligible space dependence induced by an appropriately tuned Fano-Feshbach resonance\,\cite{Fano, Feshbach, Feshbach:rev}. Although strictly valid for ultracold gases in an external potential,  for simplicity we will assume that also in the relativistic case the only inhomogeneous background quantity is the speed of sound, or equivalently, the $\rhot$ mass, while $\rho_0$ is homogeneous. 
\begin{figure}
    \includegraphics{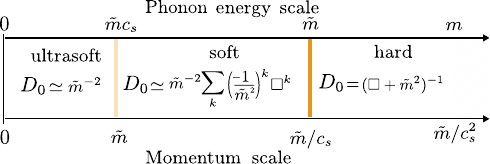}
    \caption{Schematic representation of the scale separation between ultrasoft, soft and hard scales.  Spatial momentum (bottom axis) and phonon energy (top axis)   characterizing the different scales are reported in terms of the bare boson mass, $m$, the  effective mass  of the radial field fluctuation, $\tilde{m}$,  and the adiabatic speed of sound, $c_s$,  see the main text for more details. 
    The ultrasoft  theory  is equivalent to the hydrodynamic limit, with phonons the only effective degrees of freedom, while the $\rhot$ field is static. At the soft scale the radial field fluctuations become dynamical. The hard scale is equivalent to the microscopic scale.   
    We also report the corresponding approximation of the bare radial-field propagator, $D_0$, see Eq.\,\eqref{eq:bare_propagators}.  In the ultrasoft limit it can be approximated by $1/\mt^2$, in the soft limit  we can treat it in a power expansion while for energies above $\mt$  the relativistic  propagator should be used.}
\label{fig:scheme}
\end{figure}
In principle, this also  means that one should take  into account the space modulation of the  healing length
\be\label{eq:xi}
\xi^2 \sim \frac{1}{ m g n} \sim \frac{1}{\mt^2}\,,
\ee
which  determines the shortest  wave-function variation scale for the single particle solution of the GP equation. 

For distances much larger than the healing length, corresponding to   energies well below $\mt$, the radial mode can be assumed to be static, thus the only dynamical degrees of freedom are phonons. This is known as the hydrodynamic limit. More precisely, for $c_s < 1$, we can distinguish three  energy scales, $m \sim \mt/c_s$, $\mt$ and $\mt c_s$, which  we  call  hard,  soft and  ultrasoft scales, respectively, see Fig.\,\ref{fig:scheme} for a schematic representation. We define the  corresponding momentum scales  by dividing the energy scales by the speed of sound, (we assume that this is as well the scale of the background fluid velocity), thus the   hard,  soft and ultrasoft scales correspond respectively to momenta $\mt/c_s^2$,  $\mt/c_s$ and $\mt$.     Therefore, the hydrodynamic limit is equivalent  to the ultrasoft scale, corresponding to spatial momenta much less than  $\mt$.  Close to the soft  scale, both phonons and radial oscillations are dynamical. Finally, the hard scale is  the microscopic scale, corresponding to single atom  excitations. Having a space dependent healing length implies that such scale separation may change within the system. Since we will treat the space modulation as a small perturbation, we neglect such effect, and we will refer to $\xi$ as the average value for the healing length in the system.

\section{Partition function}
\label{sec:relations}

In the Madelung representation, the partition function is given by
\be
Z[{\bm J}]= \int \mathcal{D} \rho \mathcal{D}\theta \exp \left[ i \! \int \! d^4\! x \, \left( \mathcal{L} + \!    J_\rho \rho + J_\theta \theta \right) \right]\,,
\ee
where
$\boldsymbol{J} = (J_\rho , J_\theta)$ are the external currents. By functional derivatives, we can extract the correlation functions of the fluctuations, expanding the Lagrangian density close to the stationary solution, see Eq.\,\eqref{eq:expansion}. Including terms up to the second order in the fluctuations  
\be 
\lag = \lag_0 (\rho_0, \theta_0) + \lag_2 (\rhot, \thetat) \,,  
\ee
 the  generating functional can be written as $Z = Z_0 \tilde{Z} $, where $Z_0$ is evaluated at the stationary point, while
\be \label{eq:Ztilde}
\tilde{Z}[\boldsymbol{J}]= \int \mathcal{D} \rhot \mathcal{D}\thetat \exp \left[ i \! \int \! d^4\! x \left( \mathcal{L}_2 + J_\rho \rhot + J_\theta \thetat \right) \right]\,,
\ee
contains the quadratic Lagrangian given in Eq.\,\eqref{eq:lagrangian_fluctuations}, which  can be cast as
\begin{equation}\label{eq:lagrangian_fluctuations_2}
\mathcal{L}_2 = -\frac{1}{2}  \rhot (\square + \tilde{m}^2) \rhot  + \frac{1}{2}\de_\mu \thetat \de ^\mu \thetat +  \rhot \de_L \thetat\,, \end{equation}
with
\be\label{def:longitudinal}
\de_L = \frac{V^\mu}{\rho_0} \de_\mu\,,
\ee
the longitudinal derivative, which  is an operator of dimension $2$ in natural units. We have also rescaled 
\be
\thetat \to \frac{\thetat}{\rho_0}\,,
\ee
to simplify the notation. This is equivalent to define the scaling factor in Eq.\,\eqref{eq:madelung}, $ f =\rho_0$, which is possible because, with no significant loss of generality,  we are assuming that $\rho_0$ is nonvanishing and homogeneous.  

From the above Lagrangian, one can readily see that the bare two-point  functions of the 
$\rhot$ and $\thetat$ fields can be determined from the equations
\begin{align} \label{eq:bare_propagators}
(\square_x + \tilde{m}^2 ) D_0 (x, y)  =   \delta ^4 (x-y) \quad \text{and} \quad
\square_x G_0 (x, y)  =   \delta ^4 (x-y)
\,.
\end{align}
{We can now calculate the general expressions of the two-point  functions by functional derivatives of the partition function: }
\begin{align} 
\la \rhot (x) \rhot (y)\ra & = - \left. \frac{1}{\tilde{Z}[0]} \frac{\delta ^2 \tilde{Z}[\boldsymbol{J}]}{\delta J_\rho (x) \delta J_\rho(y)} \right|_{\boldsymbol{J} = 0}\, ,\\
\la \thetat (x) \thetat (y)\ra & = - \left. \frac{1}{\tilde{Z}[0]} \frac{\delta ^2 \tilde{Z}[\boldsymbol{J}]}{\delta J_\theta (x) \delta J_\theta(y)} \right|_{\boldsymbol{J} = 0}\, ,\\
\la \rhot (x) \thetat (y)\ra & = -\left.\frac{1}{\tilde{Z}[0]}  \frac{\delta ^2 \tilde{Z}[\boldsymbol{J}]}{\delta J_\rho (x) \delta J_\theta(y)} \right|_{\boldsymbol{J} = 0}\,,
\end{align}
{which can be extracted from Eq.\,\eqref{eq:Ztilde} because $Z_0$ depends only on the stationary solution of the fields.
The integrand in Eq.\,\eqref{eq:Ztilde} is by construction gaussian in both the the radial and phonon field fluctuations, thus  we can integrate out both of them}. We begin with  the $\rhot$ field. Integrating it out is equivalent to substitute in $\tilde{Z}$ the saddle point solution of
\be 
- (\square + \tilde{m} ^2 ) \rhot + \de_L \thetat +J_\rho =0 \,,
\ee
that we can formally invert as
\be \label{eq:rhot}
\rhot (x) = D_0(x, y) [ \de_L^y \thetat (y) +J_\rho (y)] \, ,
\ee
where for the sake of notation we assume that the repeated variables are integrated. Integrating out the $\rhot$ field the partition function reads
\be
\tilde{Z}[\boldsymbol{J}]=  \tilde{Z}[0] \int \mathcal{D} \thetat \, \exp \left[ i \! \int \! d^4\! x d^4\!y\, S[\thetat, {\bm J}] \right] \,,
\ee
with $\tilde{Z}[0]$ the normalization factor and 
\be
S= S_{\thetat}+S_\text{int} + S_{J_\rho}\,,
\ee
the total action, which includes the contribution of the currents.
The term quadratic in the $\thetat$ field is
\be
S_{\thetat} = \frac{1}{2} \de_\mu^x\thetat (x) \left( \eta^{\mu\nu}\delta^4(x-y)  +\frac{V^\mu V^\nu}{\rho_0^2} D_0(x,y)\right) \de_\nu ^y\thetat(y)\,,
\ee
integrating by parts and neglecting surface terms, we can rewrite it as
\be
S_{\thetat}=-\frac{1}{2} \thetat (x) G^{-1}(x,y) \thetat(y)\,,
\ee
where the inverse propagator of the $\thetat$ field is 
\be\label{eq:inverse_G}
G^{-1}(x,y)=  \delta^4(x-y)  \square_x  -\de_L^x \de_L^y D_0(x,y)\,,
\ee
with the longitudinal derivative  defined in Eq.\,\eqref{def:longitudinal}. The full  phonon propagator,  is given by the solution of the equation
 \be\label{eq:G-1G}
G^{-1}(x,z)G(z,y)=\delta^4(x-y)\,,
\ee
thus it   takes into account the  radial field  bare propagator.

\begin{figure}
    \includegraphics[scale=1.2]{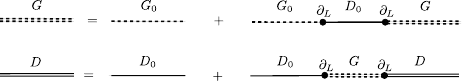}
    \caption{Feynman diagram representation of the Dyson-like equations \eqref{eq:dysonG} (top)  and~\eqref{eq:dysonD} (bottom). Thick double (solid or dashed) lines correspond to full propagators, while
    thin (solid or dashed) lines to bare propagators. The phonon propagators  are represented by dashed lines, while $\rhot$ propagators by solid lines. The vertices are the derivative operators defined in   Eq.\,\eqref{def:longitudinal}.}
    \label{fig:feynman_1}
\end{figure}

The action 
\begin{align}
S_\text{int} = \frac{1}{2}  J_\rho(x)  D_0(x,y)  \de_L^y \thetat(y) + \frac{1}{2}    \de_L^x \thetat(x) D_0(x,y) J_\rho(y)  + J_\theta(x)  \thetat(x)\,,
\end{align}
describes the interaction of the phonon with the external currents; since  the bare propagator $D_0$ is symmetric, it can be rewritten as
\begin{align}
S_\text{int} =&   J_\rho(x)  D_0(x,y)  \de_L^y \thetat(y)  + J_\theta(x)  \thetat(x)\,.
\end{align}
Finally,
\be
S_{J_\rho} = \frac{1}2 J_\rho(x) D_0(x,y) J_\rho(y)\,,
\ee
is the standard  term  quadratic in the external currents $J_\rho$. 

We can now integrate out the $\thetat$ field using the same procedure above. The equation of motion of the phonon field reads,
\be
G^{-1}(x,y) \thetat(y) = J_\theta(x) -  
\de_L^x D_0(x,z) J_\rho(z)\,, 
\ee
that we can formally invert to obtain
\be 
\thetat(x) = G(x,y) \bigg[ J_\theta(y) -  \de_L^z D_0(y,z) J_\rho(z)   \bigg]\,,
\ee
 where we have taken into account that  in general the derivative operator and the propagator $D_0(x,y)$ do not commute, because the radial mode mass can be space-time dependent. The final form of the partition function is 
\be\label{eq:ZW}
\tilde{Z}[\boldsymbol{J}]=  \tilde{Z}[0]  e^{i      \, {W}[\boldsymbol{J}] } \,, 
\ee
where
\be
W[\boldsymbol{J}]= \frac{1}{2} J_\theta G J_\theta + \frac{1}{2}J_\rho D_0 J_\rho + \frac{1}2 (\de_L D_0 J_\rho) G (\de_L D_0 J_\rho) -\frac{1}{2}J_\theta G   \de_L(D_0 J_\rho) - \frac{1}{2} (\de_L D_0 J_\rho)  G J_\theta \,,
\label{eq:wj}
\ee
is the generating functional, which depends on the full 
phonon propagators, $G(x,y)$, and on the bare radial field  propagator $D_0(x,y)$. For notation convenience we have suppressed integrals and integration variables; it is understood that any product of functions  corresponds to an integration and that derivatives act on the functions on their right side. 

By functional derivatives we  readily obtain that
\be\label{eq:corr_tt}
\la \thetat (x) \thetat (y)\ra  = -i G(x,y)\,,
\ee
is the phonon two-point function. This is however in an implicit form, because $G(x,y)$ depends on $D_0$, see Eqs.\,\eqref{eq:inverse_G} and \eqref{eq:G-1G}. In order to make such a dependence manifest, we rewrite Eq,\,\eqref{eq:G-1G}  as the Dyson-like equation
\be
\label{eq:dysonG}
 G(x,y)= G_0(x,y) + G_0(x,z) \left(\de_L^z \de_L^r D_0(z,r)\right) G(r,y)\,,
 \ee
and we will refer to  the second term on the r.h.s. as the phonon self-energy. 
The above equation is depicted in the top line of Fig.\,\ref{fig:feynman_1},  where the 
full phonon propagator correspond to the double dashed line and the bare propagator is indicated with the single dashed line.  The vertex factors are indicated with dots and correspond to the operator $\de_L$, see Eq.\,\eqref{def:longitudinal}. In the present analysis,  the vertex factor  should be counted as $\sim \mu v p$ in the momentum expansion, where  $\mu v \sim \mt$ and  $p$ is the spatial momentum.  The solid thin line in Fig.\,\ref{fig:feynman_1} correspond to  the bare radial field propagator, $D_0$, while the total two-point  functions of the radial field fluctuations, $D(x,y)$, are depicted with a double solid line. By functional derivation of Eq.\,\eqref{eq:ZW}, we find that  such propagator  satisfies the Dyson-like equation  
\be
 D(x,y)  = D_0(x,y)    + D_0(x,z)  (\de_L^z \de_L^r G(z,r))D(r,y)  \,,
\label{eq:dysonD}
\ee
where we will call the second term on the r.h.s.  the radial field self-energy. This equation  is represented  on the bottom of Fig.\,\ref{fig:feynman_1}. The system of  Eqs.\,\eqref{eq:dysonG} and\, \eqref{eq:dysonD} is the central result of the present work. It allows us to consistently compute the correlation functions at any desired order in the momentum expansion.

\section{Evaluating the two-point functions  }
\label{sec:correlators}
We now proceed to evaluate the two-point functions 
at different orders in the momentum expansion. The momentum power counting is nontrivial because one should take into account that the vertices mix phonons with radial fields and therefore the contribution of different poles should be properly taken into account. Moreover, the vertex expansion corresponds to considering higher powers of momenta,  which in some cases are compensated by the momenta in the propagators. In this section we work with arbitrary $d$ space-time dimensions,  see the Appendix\,\ref{sec:dimensiona} for a brief discussion of the dimensional reduction procedure.

\subsection{Correlation functions in the ultrasoft limit}
 We will indicate the momentum of the radial field fluctuation and of the phonon  respectively  with 
  \be
 k_\mu= (k^0, -{\bm k})\quad \text{and} \quad 
 p_\mu= (p^0, -{\bm p})\,,
 \ee
and defining $k=|{\bm k}|$ and , $p=|{\bm p}|$, we will assume that 
 $k, p \ll \mt$, with $\mt$ varying on scales much larger than the size of the system, thus it can be taken as a constant.  In the ultrasoft regime we neglect all terms of order   $k/\mt$ and $p/\mt$.

Let us first consider the phonon Dyson-like equation\,\eqref{eq:dysonG}. In the self-energy diagram one has to take into account that there are two contributions, one from the phonon pole and one from the radial field pole. The phonon-pole contribution is at  energy $p^0 \simeq c_s p \ll \mt$, and by momentum conservation $k_0=p_0$, meaning that the $\rhot$  in the phonon self-energy has a high virtuality: it is far away from its mass shell.  In this case the bare radial field propagator in the phonon self-energy diagram can be approximated as
\be D_0(x,y) = \frac{1}{\mt^{2}} \delta^d (x-y)\,,
\ee and  upon substituting the above expression in Eq.\,\eqref{eq:dysonG}, and expanding in momenta,  we have a series of terms all of the same order $1/p^2$. All of these terms must be summed and as a result  we obtain  the acoustic phonon propagator, which  in  momentum space is given by
\be\label{eq:acousic_propagator}
G_A(p_1,p_2) = \frac{1}{g_{\mu\nu} p_1^\mu p_1^\nu} \delta^d(p_1-p_2)\,,
\ee
where, apart from a conformal factor,  $g_{\mu\nu}$ is the relativistic acoustic metric. In $3+1$ dimensions it is given by
\be 
\label{eq:metric}
g_{\mu\nu}= \eta_{\mu \nu} + \left(c_s^2 - 1 \right)  u_\mu  u_\nu\,,
\ee
where  $u_\mu$ is the fluid four-velocity, see\,\cite{Bilic:1999sq, Mannarelli:2008jq, Visser:2010xv} for a discussion of the relativistic acoustic metric and for different derivations. In our picture the acoustic metric emerges from the sum of an infinite series of Feynman diagrams, all of them at the same order in the momentum expansion.

The second contribution is from the radial field pole. In this case both phonons  in Eq.\,\eqref{eq:dysonG} are off-shell, with an energy $p_0 \sim \mt$. Such contribution is suppressed by a factor $p^2/\mt^2$  and has to be discarded in the ultrasoft limit. 

Regarding the self-energy diagram of the Dyson-like equation of the $\rhot$ field in Eq.\,\eqref{eq:dysonD}, we have  two contributions, as well.  When the  propagator of the radial field is on-shell, that is  $k^0 \sim \mt$, then  in the self-energy diagram the phonon energy is $p_0 = k_0 \sim \mt$  by four-momentum conservation, thus the phonon is highly virtual. For this reason the phonon propagator  contributes with a factor $\sim 1/\tilde{m}^2$  and  it follows that the first order self-energy correction is 
of order $p^2/\tilde{m}^2$, that  is negligible  in the ultrasoft limit. The second contribution arises from the phonon pole. In this case $p_0 \sim p$ and the radial field propagators contribute with $1/\mt^4$. In the end this gives a correction of the order $p/\tilde{m}$, which should be discarded as well. Thus,  in the ultrasoft limit the radial field correlation function is not affected by the phonon propagation. These results are consistent with the fact that in the ultrasoft limit phonons propagate in the emergent acoustic metric of the system induced by  the static $\rhot$ field.  Since we treat $\tilde{m}$ as a constant we can readily evaluate 
\begin{align}
D_0(x,0) = - \int  \frac{d^d k}{(2\pi )^d}  \frac{e^{ik^\mu x_\mu}}{  k^\mu k_\mu - \mt^2 + i \epsilon}
= 
i \int \frac{d^{d-1} k}{(2\pi)^{d-1} {2 \omega_k}} e^{- i{\bm k \cdot \bm x}}\left( e^{-i \omega_k t } \Theta _H(t) + e^{i \omega_k t } \Theta _H (-t) \right)\,,
\label{eq:G2_n}\end{align}
where we used the $i\epsilon$ prescription, $\epsilon\to 0^+$,  $\Theta_H$ is the Heaviside step function and $\omega_k = \sqrt{ k^2 +\mt^2}$ is the radial mode dispersion law. The two-point correlation function is obtained taking equal times, therefore $t=0$ in the present case. In quasi $1+1$ dimensions,  see the Appendix\, \ref{sec:dimensiona},  the   two-point correlation function turns out to be
\begin{align}\label{eq:D0_us} 
 D_0(x,0) & =  i \int^{\bar k}_{- \bar k} \frac{dk}{4\pi} \frac{e^{-ik  x}}{\omega_k} \simeq \frac{i}{2 \pi} {\mathcal K}_0(|{ x}|\mt)\,, 
\end{align}
where $\bar k$ is a momentum cutoff. Since the integral is convergent we actually extended the integration domain to infinity and obtained the modified Bessel function, ${\mathcal K}_0$. 
The correlation function exponentially decays with distance, with typical length scale $1/\mt \sim \xi$, see Eq.\,\eqref{eq:xi}. This happens because taking $t=0$ and $x\neq 0$ is equivalent to  propagation outside the lightcone, which is exponentially suppressed. According to Eq.\,\eqref{eq:mt} the radial field mass  vanishes at $\mu=m$, meaning that the density-density correlation function diverges: a   typical behavior close to   a second-order phase transition.

An important aspect that emerges in the discussion of  the ultrasoft limit is that the leading corrections to the correlation functions are due to the phonon poles. This result can be generalized to any order of the vertex expansion of the  Dyson-like equations \eqref{eq:dysonG} and  \eqref{eq:dysonD}: the leading   momentum contributions arise from the poles of the phonon  propagators. We will  refer to this property as the phonon-pole dominance.

\subsection{Correlation functions in the soft limit}
In order to include the effect of the phonon fluctuations in the radial field correlation function one should consider  the soft regime, where we  include the expansion terms in $p/\mt \ll 1$. 
At this order the $\rhot$ field becomes a dynamical non-relativistic mode. For simplicity we will include only the leading corrections. 

We begin with considering the $\rhot$ two-point function. As discussed above, the dominant contribution arises from the phonon pole, meaning that both the ingoing and outgoing $\rhot$ in the leading order vertex expansion of Eq.\,\eqref{eq:dysonD} have high virtuality.  This correlation function is given by \begin{align}\label{eq:D_soft}
D(x,y) = D_0(x,y) + 
\frac{1}{\mt^4} \de_L^x \de_L^y G_A(x,y)\,, 
\end{align}
where the bare propagator in the static limit is in Eq.\,\eqref{eq:D0_us} and $G_A(x,y)$ is the acoustic phonon propagator, see Eq.\,\eqref{eq:acousic_propagator}. The 
$G_A(x,y)$ and not the full phonon propagator appears in Eq.\,\eqref{eq:D_soft} because corrections to $G(x,y) $ that arise in the soft limit are by construction  subleading in $p/\mt$.    Using Eq.\,\eqref{eq:corr_tt}, we find that there is  a simple relation between the correlation functions 
\be 
\la \rhot (x) \rhot (y) \ra = -iD_0(x,y) + \frac{1}{\mt^4}  \de_L^x \de_L^y \la\thetat (x)  \thetat (y) \ra \,,
\ee
moreover we have that
\be
\la \thetat(x) \rhot (y) \ra = - i \frac{1}{\mt^2} \de_L^y G_A(x,y) = \frac{\de_L^y}{\mt^2} \la  \thetat (x)  \thetat (y) \ra \,,
\ee
 where in both cases it is understood that the phonon correlation function is evaluated in the ultrasoft limit.
 In  general,   these two-point correlations at a given order in momenta  can be obtained starting from $G(x,y)$ evaluated at the previous order of momenta using  appropriate derivatives.
 
It remains to evaluate $G(x,y)$ at this order, which however would only be useful to evaluate the $\rhot$ two-point  correlation functions at the next order in the momentum expansion.  We shall  provide the explicit results elsewhere,   here we only notice that all the leading order contributions that arise from  the phonon poles have already been summed in $G_A$. It remains to evaluate the contribution of the 
$\rhot$ poles. To this end, it is sufficient to evaluate the leading order term in the vertex expansion of Eq.\,\eqref{eq:dysonG}.


\subsection{Perturbative methods for an inhomogeneous background}
\label{sec:beyond_LDA}
So far we have assumed that the mass of the $\rhot$ field is a smooth function of the coordinates. In this section we consider the case of  a non negligible space-time dependence. We  present an  approach  based on a perturbative expansion that takes into account that an inhomogeneous medium is formally equivalent to a   space-time dependent external field, see for instance~\cite{abrikosov2012methods}. The main difference with homogeneous systems discussed above, is that now the propagators are not only a function of  $x-y$, but will  also depend on $x+y$. 

Starting from the Lagrangian density in Eq.\,\eqref{eq:lagrangian_fluctuations_2}, we can write the inverse propagator of the $\rhot$ field as
\be
\label{eq:dinv}
D_0^{-1}(x,y) \to D_0^{-1}(x,y) + \epsilon D_{0I}^{-1}(x,y)=\left( D_0^{-1}(x) + \epsilon D_{0I}^{-1}(x) \right) \delta^d(x-y)\,,
\ee
where $\epsilon$ is a perturbative parameter, $D_{0I}^{-1}$ is equivalent to an external field and the last equality holds because  we restrict ourselves to local interactions.
In momentum space,   the effect of the $\epsilon$ correction is that
\be\label{Eq:Dyson}
D_0(p_1,p_2) \to D_0(p_1) \delta ^d(p_1-p_2) - \epsilon D_0(p_1)D_{0I}^{-1}(p_1,p_2) D_0(p_2)\,,
\ee
and therefore 
\be\label{eq:D_p1p2}
D_{0I}(p_1,p_2) = -\epsilon D_0(p_1) D_{0I}^{-1}(p_1,p_2)  D_0(p_2)\,, 
\ee
which is not proportional to $\delta^d(p_1-p_2)$, because in an inhomogeneous system the momentum is not conserved. One can formally restore momentum conservation  viewing the  effect of the inhomogeneous background as originating from an external field, having a momentum that exactly compensates the momentum variation of the $\rhot$ field.  
Finally, 
\be \label{eq:D1xy}
D_{0I}(x,y) = \int \frac{d^d p_1}{(2\pi)^d}\frac{d^d p_2}{(2\pi)^d} e^{i p_1 x}e^{-i p_2 y} D_{0I}(p_1,p_2)\,,
\ee
is the correction to the propagator in coordinate space.
To recap, the procedure is as follows:  starting from $D_{0I}^{-1}(x)$  determine $D_{0I}^{-1}(p_2-p_1)$, plug it in Eq.~\eqref{Eq:Dyson} and then obtain the correction to the propagator in coordinate space by Eq.~\eqref{eq:D1xy}. The bare $\rhot$ field propagator is now modified as follows
\be
\label{eq:dnew}
D_0(x,y) \to \hat D_0(x,y) =D_0(x,y) + \epsilon D_{0I}(x,y)\,,
\ee
which has to be plugged in the Dyson-like Eqs.\,\eqref{eq:dysonG} and \eqref{eq:dysonD} to determine the variation of the full two-point correlation functions. For example, at the leading order in $\epsilon$  we have that
\begin{align}
\hat G(x,y)  = G(x,y) + \epsilon G_I(x,y)\quad \text{and} \quad
\hat D(x,y)=  D(x,y) + \epsilon D_I(x,y)\,,
\end{align}
with
\begin{align}
G_I(x,y)  = &   G_0(x,z) \left(\de_L^z \de_L^r D_{0I}(z,r)\right) G(r,y)\,,  \\
 D_I(x,y)= &  D_{0I}(x,y) +  D_{0I}(x,z)(\de_L^z \de_L^r G(z,r))D(r,y) +  D_0(x,z)  (\de_L^z \de_L^r G(z,r))D_{0I}(r,y)
\nonumber \\ & +  D_0(x,z)  (\de_L^z \de_L^r G_I(z,r))D(r,y)\,,
\end{align}
where the last term in the second equation arises from a modification of the phonon correlation function and it is therefore subleading in the vertex expansion. It is important to notice that in the momentum expansion  one should consider  Eq.\,\eqref{eq:D_p1p2}, meaning that for the momentum power counting, the perturbation amounts to two unperturbed radial field propagators.  
In any case, for a given external modulation of the speed of sound it is now possible to evaluate the effect on the two-point correlation functions at any desired order in the momentum expansion. 

We are now in a position to undertake the study of  any  
inhomogeneous superfluid system employing a double expansion: in momenta and in $\epsilon$. This means that we can determine with arbitrary accuracy
the two-point correlation functions of inhomogeneous superfluids. Notice that this can be done recursively: because of the phonon-pole dominance, the phonon propagator at a given order in momentum expansion determines the variation of the density-density correlation function at the next order in momenta.


\section{Conclusions}
\label{sec:conclusions}
{We have developed a path integral method to compute the correlation functions of low-energy excitations in a weakly interacting boson superfluids with background inhomogeneities.

Starting from the partition function in the Madelung representation, we derived an effective action up to quadratic order. This allowed us to obtain the generating functional by integrating out both radial and phonon fluctuations. Our method provides a simple and systematic way to compute two-point correlation functions using functional derivatives, while keeping the physics transparent with respect to the conceptual map in Fig.~\ref{fig:scheme} for the different and relevant momentum/energy scales dictated by the boson mass and the effective mass of the radial fluctuations. The full correlation functions for the radial and phonon fields are then expressed as two Dyson-like equations, see Fig.\,\ref{fig:feynman_1} and  Eqs.\,\eqref{eq:dysonG} and\, \eqref{eq:dysonD}, which can be then expanded in powers of moments over the effective mass of the radial field. Besides its generality, our approach can reveal to be especially useful in building up effective field theories in curved space-time, one of the core subjects of Renaud Parentani's contributions to the scientific community\,\cite{Arteaga:2005ss, Macher:2009tw, Finazzi_2012, Michel_2016}.

We analyzed these correlators in the two relevant regimes, corresponding to the ultrasoft and soft energy scales. In the ultrasoft limit the density-density correlation function is insensitive to the phonon field, thus it shows exponential decay, as expected from propagation outside the lightcone. The typical correlation length is in this case the inverse radial field mass, of the order of the superfluid healing length. When the effective mass of the radial field approaches zero, the correlation function diverges, signaling the presence of a second-order phase transition. We also observed  the phonon-pole dominance, meaning that  at each order of the vertex expansion of  the Dyson-like equations, the leading contribution  comes from the phonon propagator pole. This implies that we can  properly organize the momentum expansion in a recursive way:  the phonon two-point function evaluated at the $n$th order of the momentum expansion allows to  evaluate the density-density two-point function at the $(n+1)$th order in momenta. In turn, this can then be used to evaluate the  phonon two-point function at the $(n+1)$th order. 

Finally, we introduced a perturbative method to calculate the correlation functions in an inhomogeneous background. Since the method is perturbative, it allows for a recursive computation of the two-point correlation functions with arbitrary precision, making it suitable for describing realistic and spatially varying superfluid configurations. For these reasons the present study paves the way for the systematic discussion of the density-density correlation functions in systems 
having  a modulated sound speed that makes the fluid flow transonic,  see~\cite{biondi2025effectivefieldtheoriesanaloguegravity} for a preliminary study. To this end,  one could consider a stationary fluid flowing at the constant velocity, $\bm u$,  with a  speed of sound having a small space dependence, proportional to $\epsilon$, such that in the supersonic region it is less then $|\bm u|$, while in the subsonic  region is larger than    $|\bm u|$. The two regions are separated by the sonic horizon, where the speed of sound equals the fluid velocity.  The presented method should allow to perturbatively determine the effect of the sonic horizon on the two-point correlation functions. In particular, one could  determine the imprint of the Hawking radiation,  which is expected to be emitted close to the acoustic horizon, on the density-density correlation function.

This work is limited to the quadratic (non-interacting) case, but it lays the groundwork for future studies that include interactions. 
The procedure is in principle  straightforward: one should simply properly include the interaction vertices in the momentum power counting, while at the same time accounting for the relevant Ward identities. In this way one could as well determine the effect of interactions on the Hawking-like radiation finding  how it  eventually  modifies the spectrum of the emitted phonons. This work can in principle be generalized also to systematic accounting of inhomogeneities, by reverting to the formal apparatus of time-dependent density-functional theory in connection with quantum hydrodynamics ~\cite{VUC1997} in the relevant case of superfluids~\cite{Chiofalo01,Chiofalo1998}.

\appendix
\section{Dimensional reduction}
\label{sec:dimensiona}
In many experiments, the system is effectively $1+1$ dimensional: the ultracold atoms are in a trap elongated in one direction, while the transverse extension is much smaller. For this reason  it turns useful to express the correlation functions in 
quasi $1+1$ dimensions. By this we mean  a physical system with $3+1$ dimensions elongated, say along the $x-$direction,   with  transverse dimensions much smaller than the longitudinal one. To treat a $3+1$ dimensional system with  transverse size, $\ell_\perp \lesssim \xi$, we have to properly quantize it.  Let us   consider a generic  mode with mass $m$. Taking the transverse plane orthogonal to the $x-$direction,    the  free-particle dispersion law is given by 
\be
\omega_p = \sqrt{p_z^2 + p_x^2 + p_y^2+m^2}\,,
\ee
where  the properly quantized transverse momenta are
\be p_y = n_y \frac{2 \pi}{\ell_\perp}\quad \text{and}  \quad p_z = n_z \frac{2 \pi}{\ell_\perp}\,,
\ee
with $n_y$ and $n_z$ integers. The integral in momentum space now reads
\be
\int d^3 \!p \rightarrow  \left(\frac{2 \pi}{\ell_\perp} \right)^2 \sum_{n_y}\sum_{n_z} \int d\! p_x\,,
\ee
and thus for vanishing $\ell_\perp$ it diverges unless $n_y = n_z =0$, corresponding to  a mode with {\it frozen} transverse dynamics.  Assuming that the system is in a trap that freezes the dynamics in the transverse plane, we can  express the integral in momentum space for a quasi $1+1$ dimensional system as
\be
\int d^3\! p  \rightarrow  \left(\frac{2 \pi}{\ell_\perp} \right)^2  \int d^3 \!p\,\, \delta(p_y) \delta(p_z)\,.
\ee
Let us now consider the correlation function of the $\rhot$ field. From the above reasoning it follows that  the relation between the correlation function in  $3+1$ dimensions in an elongated trap and the one  in $1+1$ dimensions is simply
\be
 \left.\la \rhot (0,{\bm x}) \rhot(0) \ra\right|_\text{ $3+1$ D}    =    \frac{1}{\ell_\perp^2}   \left. \la \rhot (0,{\bm x}) \rhot(0) \ra \right|_\text{ $1+1$ D}\,,
\ee
where $\bm x$ is  along the longitudinal  direction of the trap.  When discussing quasi $1+1$ dimensional system in the main text, we are actually considering  a $3+1$ dimensional system with  frozen dynamics in the transverse plane.

\bibliographystyle{crunsrt}

\bibliography{methods.bib}

\end{document}